# Metal Mesh IR Filter for wSMA

Chao-Te Li*[a], C.-Y.E. Tong[b], Ming-Jye Wang[a], Tse-Jun Chen[a], Yen-Pin Chang[a],
Sheng-Feng Yen[a], Jen-Chieh Cheng[a], Wei-Chun Lu[a], Yen-Ru Huang[a]
[a]Academia Sinica, Institute of Astrophysics and Astronomy, Taipei 106, Taiwan;
[b]Harvard-Smithsonian Center for Astrophysics, Cambridge, MA 02138, USA

## ABSTRACT

Since the start of full science operations from 2004, the Submillimeter Array has been implementing plans to expand IF bandwidths and upgrade receivers and cryostats. Metal mesh low-pass filters were designed to block infrared (IR) radiation to reduce the thermal load on the cryostats. Filters were fabricated on a quartz wafer through photolithography and coated with anti-reflection (AR) material. The filters were tested from 200 to 400 GHz to verify their passband performances. The measurement results were found to be in good agreement with EM simulation results. They were tested in the far-infrared (FIR) frequency range to verify out-of-band rejection. The IR reflectivity was found to be approximately 70%, which corresponded to the percentage of the area blocked by metal.

## I. INTRODUCTION

Submillimeter Array (SMA) has been in operation since its dedication in 2004 and is on track to expand its IF bandwidths (wideband Submillimeter Array, wSMA) [1] and upgrade receivers and cryostats. In the new cryostat, the plan is to install low-pass infrared (IR) filters on the 65 K radiation shield to reduce the thermal load. Although a plain quartz wafer coatings can also be used to absorb IR [2], However, limited by thermal conduction between the filter and the radiation shield, the filter may heat up and re-emit at an elevated temperature to the 4 K stage. To minimize these effects, a low-pass filter that can reflect IR radiation is preferred. Such thermal filters have been used with large aperture detector arrays to minimize background radiation [3].

Ade *et al.* [4] provided a comprehensive review of metal mesh filters. Metal mesh filters are characterized as inductive, capacitive, or resonant grids, and provide high-pass, low-pass, and bandpass properties, respectively. We designed a capacitive metal mesh filter to block IR radiation of approximately 10 µm or 30 THz, where the maximum intensity of radiation emitted by a blackbody at 300 K is. In the following sections, the design, fabrication, and testing of the IR filter are described.

## II. IR FILTER DESIGN

The design began with a layer of capacitive metal mesh. Because of the periodicity of the grid, a unit cell with symmetric walls [5] was used during 3D EM simulations. In a low-pass grid, the grid period corresponds to a frequency where cutoff occurs. A grid period of 40 µm, corresponding to a cutoff frequency of approximately 7.5 THz, was chosen initially. However, the bandpass loss around 500 GHz was close to 0.4 dB. Another trial is to use a grid period of 12 µm with a cutoff frequency of approximately 25 THz. However, to achieve a blocking area percentage of more than 70%, the gap between gold patches would be less than 2 µm, which may not be easy for the lift-off process used in fabrication. Therefore, a grid period of 20 µm with gaps of 3 µm was adopted considering the in-band transmission, out-of-band rejection, and fabrication. Fig. 1 shows the model of a unit cell with the capacitive grid in the 3D EM simulator HFSS [6]; the simulated responses are shown in Fig. 2. The cutoff frequency was approximately 15 THz, where the grid period is close to the free-space wavelength. Diffraction occurred above 15 THz, when the wavelength was less than the grid period.

*ctli@asiaa.sinica.edu.tw

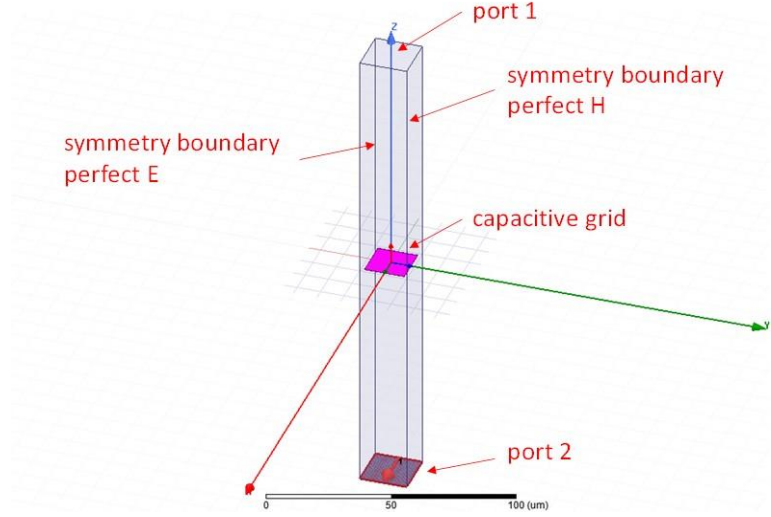

Fig. 1. 3D EM model of a capacitive grid in a unit cell. The unit cell is 20 µm × 20 µm × 200 µm with symmetrical boundaries. The capacitive grid is a 0.1-µm-thick gold patch of 17 µm × 17 µm.

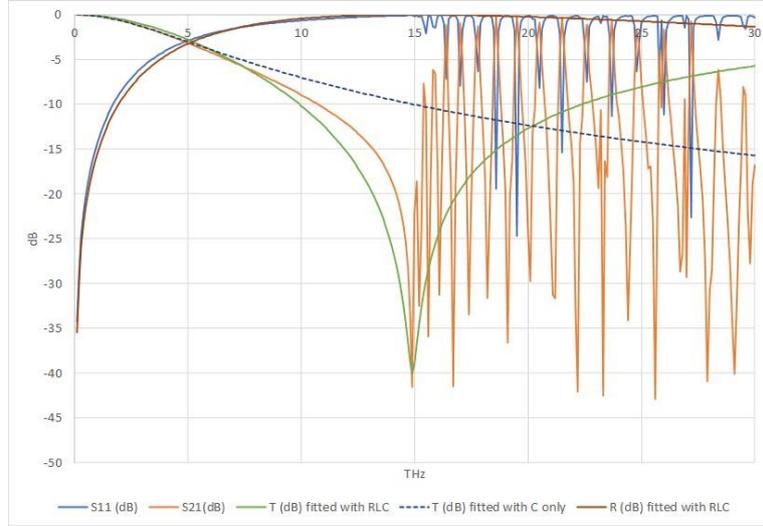

Fig. 2. Simulations of a capacitive grid of 17 µm × 17 µm with a grid period of 20 µm (orange line: transmission, blue line: reflection). The dotted line represents the fitting with a capacitor alone. Fitting with the modified equivalent circuit is also shown (green line: transmission, brown line: reflection).

Modeling provides us with insights into the characteristics of the metal mesh and is useful for multilayer filter design. An equivalent circuit with a shunt capacitor, as shown in Fig. 3, was used to model the capacitive grid. The transmission is expressed as

$$|S_{21}|^2 = |T|^2 = \frac{1}{1+(\omega C)^2} \tag{1}$$

normalized to the free-space wave impedance $\eta_0$ of 377 Ω. By fitting the transmission at lower frequencies, the capacitance was estimated to be approximately $3.1 \times 10^{-14}$ F. A modified equivalent circuit was suggested [7] to include an inductor and a resistor to account for self-resonance and loss. For the modified equivalent circuit, the transmission T and reflection $\Gamma$ are expressed as

$$|S_{21}|^2 = |T|^2 = \frac{\left(\omega L - \frac{1}{\omega C}\right)^2 + R^2}{\left(\omega L - \frac{1}{\omega C}\right)^2 + (R+1)^2} \qquad (2)$$

and

$$|S_{11}|^2 = |\Gamma|^2 = \frac{1}{(\omega L - \frac{1}{\omega C})^2 + (R+1)^2}. \qquad (3)$$

For a capacitance of approximately $3.1 \times 10^{-14}$ F and resonant frequency of 1.5 THz, the inductance was estimated to be $3.7 \times 10^{-15}$ H.

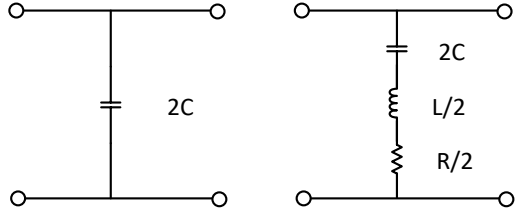

Fig. 3.  [Left] Single-element equivalent circuit for a capacitive grid. [Right] Modified equivalent circuit including inductance and resistance

A slab of a quartz substrate was added to the model, along with two quarter-wavelength AR coatings at the top and bottom, as shown in Fig. 4. Crystalline quartz with a dielectric constant of 4.45 [8] was chosen as the substrate. Its initial thickness was 860 µm, which corresponded to approximately two wavelengths at 345 GHz. Low-density polyethylene (LDPE), which has a refractive index of approximately 1.52 [8], was used as the AR material. Transmission was optimized between 210 and 360 GHz by varying the thicknesses of the quartz wafer and AR coatings. The optimal parameters of the IR filter are listed in Table I. The simulated spectral responses are shown in Fig. 5.

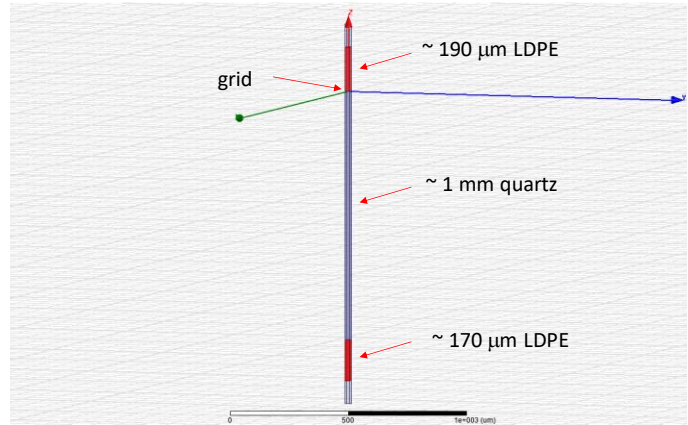

Fig. 4.  3D EM model of the capacitive grid on a quartz wafer (gray section) with LDPE AR coatings on top and bottom (red sections)

TABLE I

wSMA IR FILTER PARAMETERS

| Frequency range | 210 GHz – 360 GHz | |
|---|---|---|
| Passband loss | < 2% | |
| IR blockage | > 70% | |
| Substrate | 4" z-cut crystalline quartz | |
| | dielectric constant | 4.45 |
| | thickness | 1078 μm ± 25 μm |
| Capacitive grid | grid period | 20 μm |
| | gold patch size | 17 μm x 17 μm |
| | metal thickness | 0.2 μm |
| LDPE AR coating | refractive index | 1.52 |
| | thickness (grid side) | 191 μm ± 10 μm |
| | thickness (the other side) | 174 μm ± 10 μm |

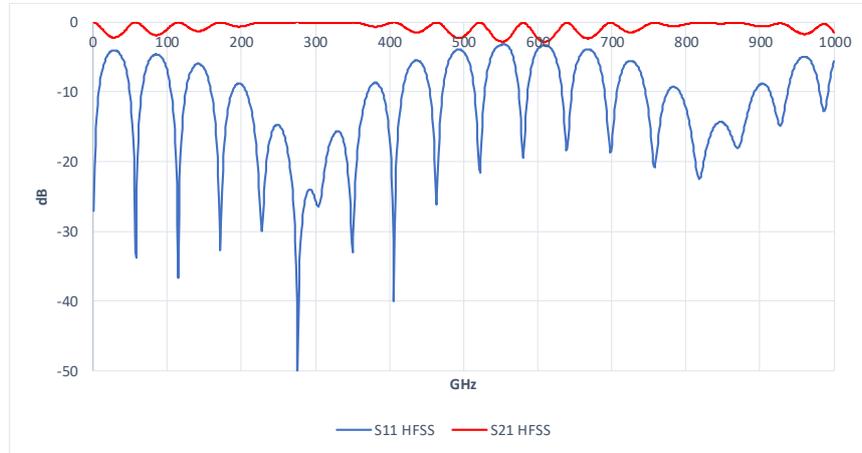

Fig. **5**.  3D EM simulations of the IR filter.

## III.  FABRICATION

A 4" z-cut both side polished quartz wafer was used to fabricate the filter. The thickness of the filter was specified to be 1.078 mm ± 25 μm. An approximately 200-nm-thick gold layer was deposited via e-beam evaporation with a thin (approximately 20 nm) layer of Ti underneath. The lift-off process was followed to form the capacitive grid. Few images taken after the processing are shown in Fig. 6.

After the capacitive grid was processed, LDPE films were attached on both sides of the filter as AR coating. The LDPE films were thermally bonded [2] to the quartz wafer at 0.25 psi in an oven at 180 °C for 6 hours. Because the LDPE films in our inventory are of only a few thicknesses, two films were combined to produce a single-layer coating. The final thickness of the LDPE film was 186 µm on the grid side and  153 µm on the other side; both these values were lower than the design values. Images of the IR filter with the AR coatings are shown in Fig. 7.

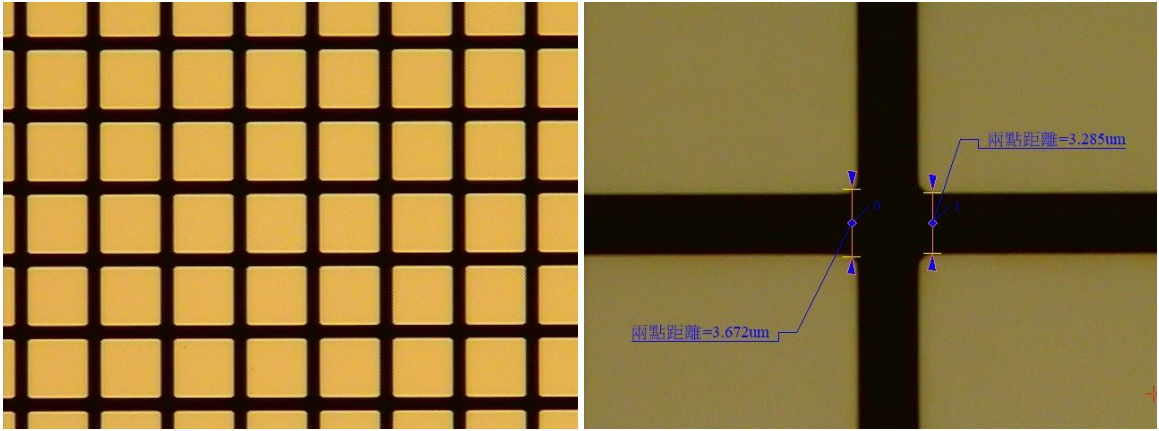

Fig. 6. Images of gold capacitive grids on a quartz wafer.

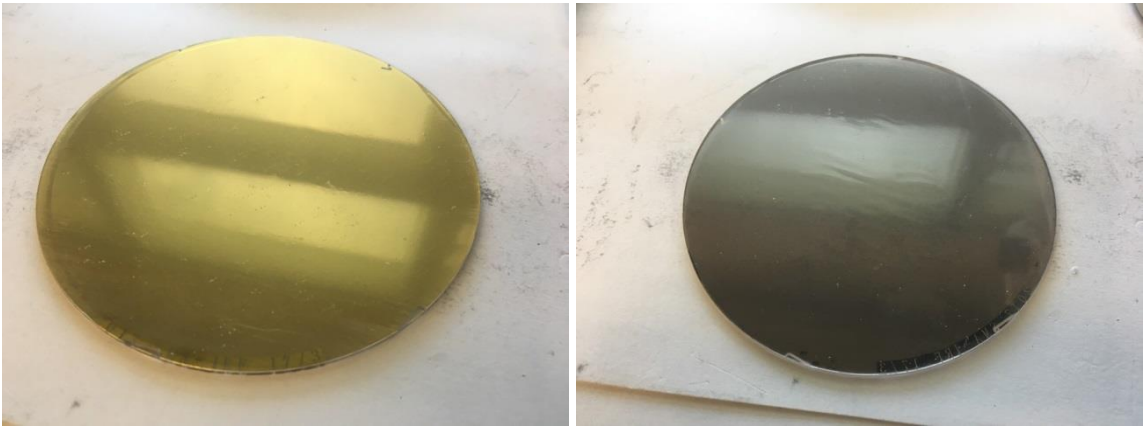

Fig. 7. Images of the front and back sides of the 4" IR filter with LDPE AR coating.

## IV. LAB TESTING

The IR filter was tested in lab at room temperature to measure the transmission and reflection within the passband. Far-infrared (FIR) Fourier transform spectroscopy (FTS) was also used to verify the out-of-band behavior up to 30 THz.

*A. In-Band Test*

A scalar system was set up to measure the passband properties of the IR filter from 200 to 400 GHz. The transmission measurement setup is shown in Fig. 8. Tone signals up to 20 GHz were provided with a signal generator and then frequency multiplied. The test tone was transmitted by a horn and propagated using two lenses that formed a Gaussian beam telescope [9] to the receiving horn. The IR filter was placed around the beam waist between the two lenses. The received power was measured using a millimeter-wave power meter. Measurements were performed with and without the IR filter. A comparison of the two sets of results determines the transmission of the IR filter, as shown in Fig. 9.

Reflection measurements were also performed using the setup shown in Fig. 10. The reflected power was measured using either the IR filter or a reference plate. A quartz wafer deposited with a 200-nm-thick gold layer was used as the reference plate. The reflection of the IR filter was derived and the results are shown in Fig. 11. The reflection pattern shifted to higher frequencies compared with that obtained in the 3D EM simulations. This was attributable to the thinner AR coatings used. Another filter with slightly thicker AR coatings (196 µm on the grid side, and 159 µm on the other side) was tested using a quasi-optical vector network analyzer (QO-VNA) [10]. The measurement results obtained with this filter were in better agreement with the simulation results, as seen in Fig. 11.

The improved method of intersecting lines [11] was used to measure the insertion loss of the mesh filter. The mesh filter was placed in front of the window of an SMA receiver operating at LO 264 GHz. The difference in intercept was 3.9 (±0.5) K, corresponding to a loss of approximately 1.2% (±0.2%) at room temperature, as shown in Fig. 12.

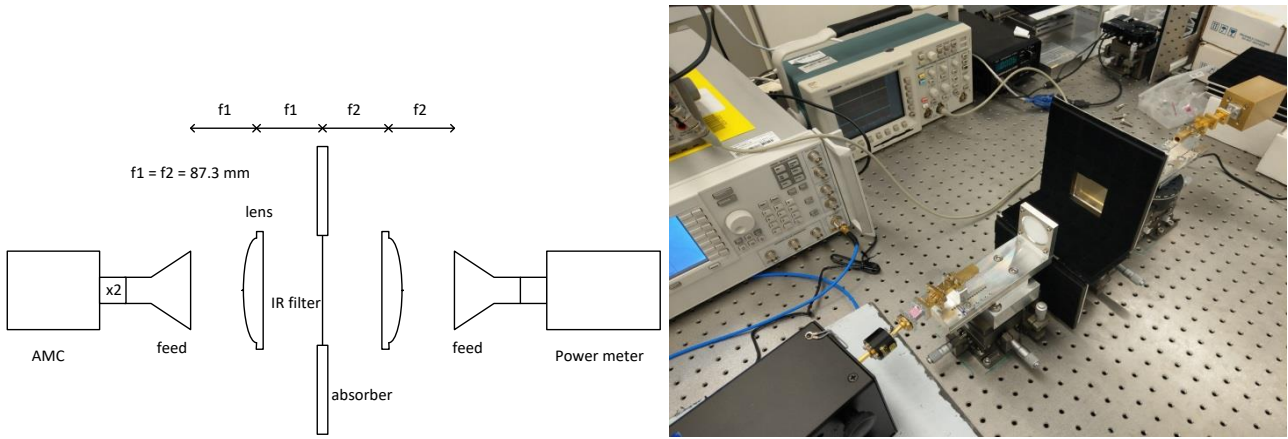

Fig. 8. Schematic and image of the in-band transmission measurement setup.

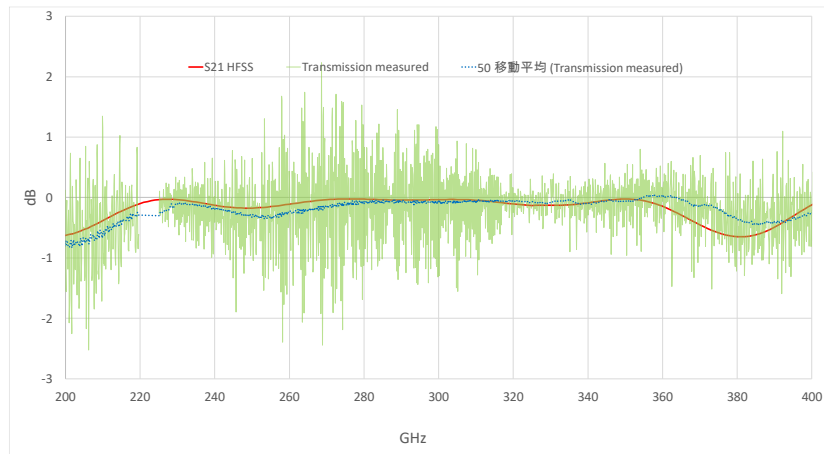

Fig. 9. Transmission of the IR filter within the passband. Measurement data were scattered mainly because of the multi-reflection between components. The data were shifted to higher frequencies compared with the 3D EM simulations (solid line).

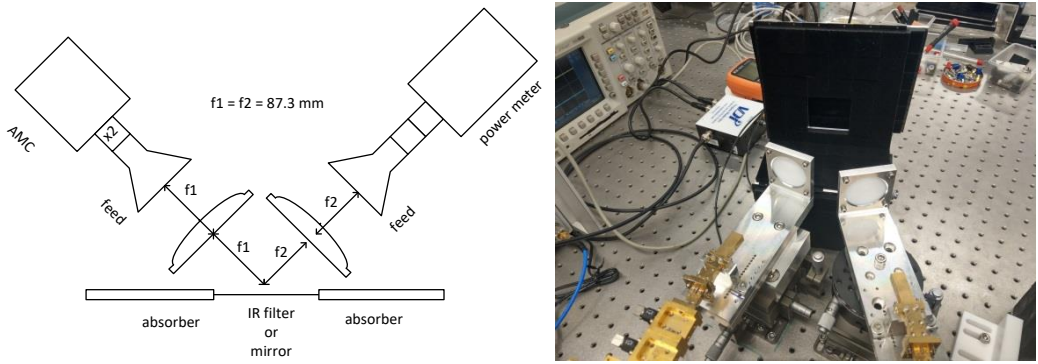

Fig. 10.  Schematic and image of the scalar reflection measurement setup.

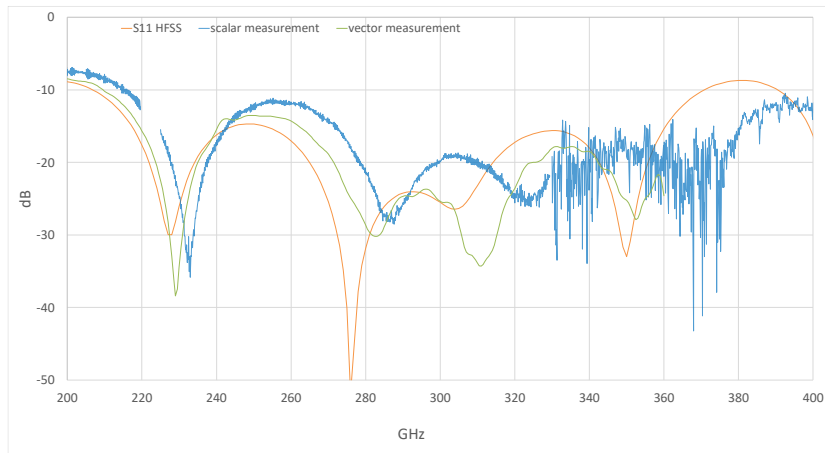

Fig. 11.  Measurement results (blue line: scalar, green line: vector) of the reflection of the two IR filters along with 3D EM simulation results (orange line).

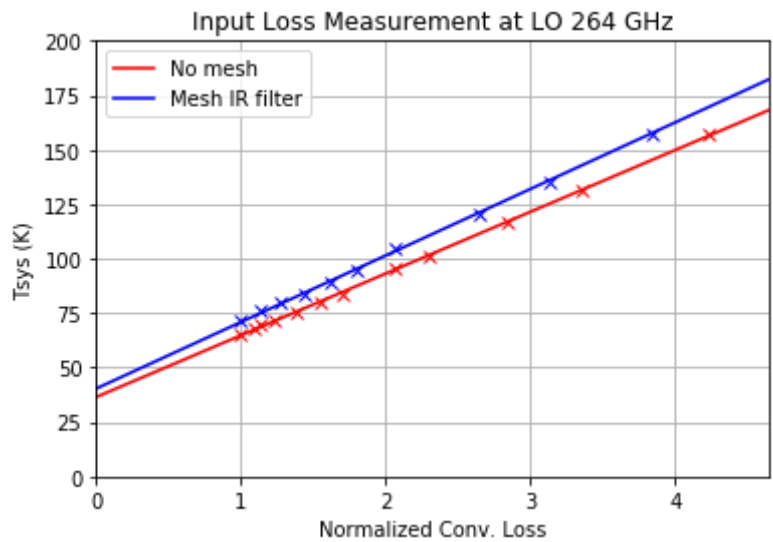

Fig. 12.  Improved method of intersecting lines to measure the added optics loss due to the mesh filter.

## B. FTS Testing

FIR FTS was used to measure the out-of-band properties of the IR filter. The transmissions of a quartz wafer and the IR filter were measured, and the results are shown in Fig. 13. The measurements were performed in two frequency ranges (one up to 10 THz, and the other up to 30 THz). The results show that the IR filter blocked FIR radiation above 5 THz, where the quartz wafer still has some transmission. The reflection measurements show that the IR filter achieved approximately 70% reflectivity in the FIR frequency range; this value corresponds to the percentage of area blocked by the capacitive grid ($17^2/20^2 = 72\%$), as shown in Fig. 14. Reflectivity of the quartz wafer was approximately 20%.

Given the reflectivity of the IR filter and quartz wafer, we can estimate the radiation heat transfer between the quartz vacuum window and the IR filter by using the radiation network method [12]. Based on Equation (4) and the parameters listed in Table II, the radiation heat transfer from the vacuum window to the IR filter is estimated to be approximately 1.3 W, lower than 2.5 W when a plain quartz wafer is in place.

$$\dot{Q}_{12} = \sigma(T_1^4 - T_2^4)\left(\frac{1-\epsilon_1}{A_1\epsilon_1} + \frac{1}{A_1 F_{12}} + \frac{1-\epsilon_2}{A_2\epsilon_2}\right)^{-1} \qquad (4)$$

TABLE II

VACUUM WINDOW AND IR FILTER RADIATION HEAT TRANSFER PARAMETERS

| Vacuum window | Material | Quartz |
|---|---|---|
|  | Temperature $T_1$ (K) | 300 |
|  | Area $A_1$ (m$^2$) | 0.015 |
|  | Emissivity $\varepsilon_1$ | 0.8 |
| IR filter | Metal mesh on quartz |  |
|  | Temperature $T_2$ (K) | 65 |
|  | Area $A_2$ (m$^2$) | 0.008 |
|  | Emissivity $\varepsilon_2$ | 0.3 |
| Distance between vacuum window and IR filter | 0.09 m |  |
| View factor | $F_{1\to 2}$ | 0.17 |

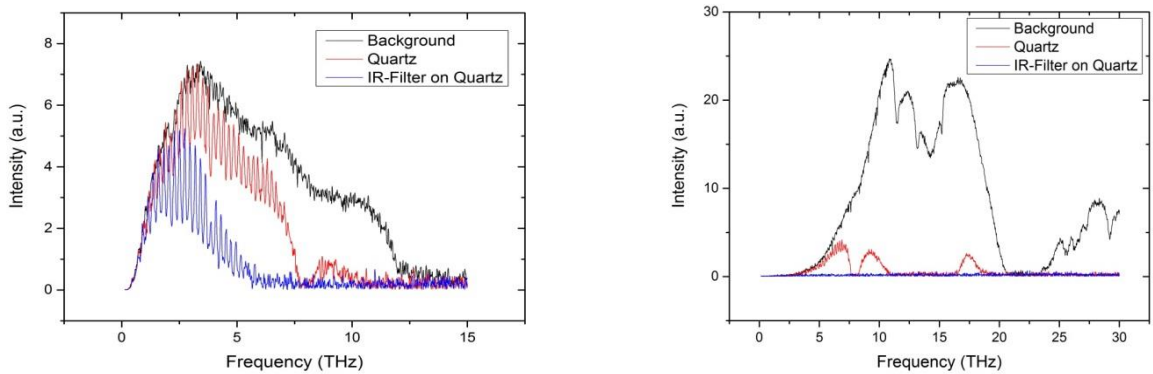

Fig. 13. Measurement results of the transmission of the IR filter and a quartz wafer in two frequency ranges using FIR FTS. Background curves provide the transmission without any sample in the optical path.

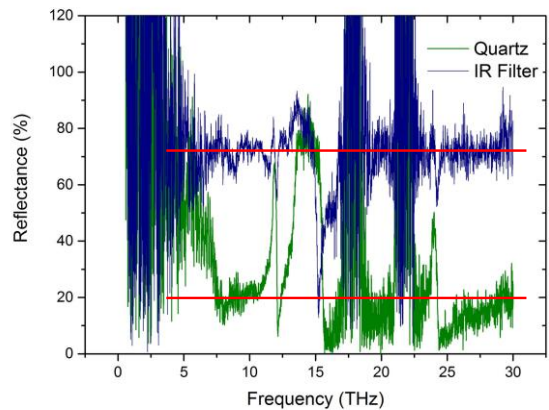

Fig. 14. Reflection measurement results of the IR filter and a quartz wafer using FIR FTS.

## V. SUMMARY

We designed, fabricated, and tested a metal mesh IR filter for wSMA to reduce the thermal load on cryostats. The in-band transmission and reflection measurement results were in good agreement with the EM simulations. FTS measurements indicated that the IR filter achieved a 70% reflectivity in the FIR frequency range, which corresponded to the percentage of area blocked by the capacitive grid.